\documentclass[pdflatex,sn-nature]{sn-jnl}%

\usepackage{graphicx}%
\usepackage{amsmath,amssymb,amsfonts}%
\usepackage{booktabs}%
\usepackage{xcolor}%
\usepackage{etoolbox}%
\makeatletter
\patchcmd{\@maketitle}{Corresponding author(s). E-mail(s):}{Corresponding author. E-mail:}{}{\PackageWarning{manuscript}{Corresponding-author patch failed}}
\patchcmd{\email}{\textcolor{blue}{#1};\ }{\textcolor{blue}{#1}}{}{\PackageWarning{manuscript}{E-mail patch failed}}
\makeatother

\begin{document}

\title[Mimicking the zero-field Josephson diode effect]{Mimicking the zero-field Josephson diode effect with remanent magnetization and self-field}

\author*[1,2]{\fnm{Nathan} \sur{Satchell}}\email{satchell@txstate.edu}
\author[2]{\fnm{Arjun} \sur{Sapkota}}
\author[2]{\fnm{Pukar} \sur{Sedai}}
\author[1]{\fnm{Kurt} \sur{Lorenzen}}
\author[1]{\fnm{Juan} \sur{Gomez}}

\affil[1]{\orgdiv{Department of Physics}, \orgname{Texas State University}, \orgaddress{\city{San Marcos}, \state{Texas}, \postcode{78666}, \country{USA}}}
\affil[2]{\orgdiv{Materials Science, Engineering and Commercialization Program}, \orgname{Texas State University}, \orgaddress{\city{San Marcos}, \state{Texas}, \postcode{78666}, \country{USA}}}

\abstract{A Josephson junction whose critical current is larger in one direction than in the other is a diode.
With a magnetized barrier, a junction can be a diode at zero applied field, and this is commonly attributed to an intrinsic Josephson diode effect, in which the magnetic barrier breaks time-reversal symmetry.
Here we show that the same signatures arise from two ordinary ingredients: the magnetic field of the bias current in the electrodes and the remanent magnetization of the barrier.
In cross-type Nb/Pd\textsubscript{89}Ni\textsubscript{11}/Nb junctions, the self-field displaces the interference patterns of the two current polarities in opposite directions.
An in-plane remanent magnetization translates the pattern and brings the resulting asymmetry to zero field, where the junction mimics a nonvolatile diode with an asymmetry of 10 to 12\% and a polarity set by the magnetization direction.
Reversing the current or the field cannot distinguish this mimic from an intrinsic effect, but the wiring can: moving the current lead to the other arm of the electrode across the field reverses the self-field diode, in applied field and at zero field alike, whereas an intrinsic diode would be unchanged.
This gives a simple dc test for Josephson diodes in the current-perpendicular-to-plane geometry.}

\maketitle

\section{Introduction}\label{sec:intro}

A superconducting diode carries a larger supercurrent in one direction than in the other, so that over a range of bias currents it is superconducting for one polarity and resistive for the other.
Since its observation in a noncentrosymmetric [Nb/V/Ta]$_n$ superlattice~\cite{Ando2020}, the superconducting diode effect has been reported in a wide range of superconducting films and Josephson junctions~\cite{Nadeem2023,Ma2025}.
In a Josephson junction it appears as a difference between the critical currents of the two polarities, $I_c^{+} \neq I_c^{-}$, and is known as the Josephson diode effect (JDE).
A nonreciprocal critical current requires broken symmetry, either of the junction itself (an intrinsic JDE) or of the device around it, where the bias current acts back on the junction (an extrinsic JDE)~\cite{Ma2025,Wang2025}.

In the intrinsic JDE the junction barrier itself breaks inversion and time-reversal symmetry, the latter through magnetism in the barrier or through an applied magnetic field~\cite{Zhang2022}.
A barrier that combines spin-orbit coupling with an exchange field gives the junction a nonsinusoidal current-phase relation whose ground state is shifted to a finite phase difference $\varphi_0$, the $\varphi_0$ junction~\cite{Buzdin2008}, which is one route to a nonreciprocal critical current~\cite{Zhang2022}.
Finite-momentum Cooper pairing is another~\cite{Davydova2022}.
Experiments in an applied magnetic field have realized the intrinsic JDE in InAs quantum-well junctions with strong spin-orbit coupling~\cite{Baumgartner2022}, in junctions of the Dirac semimetal NiTe$_2$~\cite{Pal2022}, and in interferometers that measure $\varphi_0$ and the diode effect on the same junction and show that the two are linked~\cite{Reinhardt2024}.
Removing the applied field requires time-reversal symmetry to be broken within the junction barrier itself.
Zero-field Josephson diodes have been reported with the van der Waals barriers Nb$_3$Br$_8$~\cite{Wu2022}, multiferroic NiI$_2$~\cite{Yang2025} and ferromagnetic Fe$_3$GeTe$_2$~\cite{Hu2026}, with Pt and other metal barriers proximity-magnetized by the ferrimagnetic insulator Y$_3$Fe$_5$O$_{12}$~\cite{Jeon2022,Jeon2026}, in InAs nanowires with shells of the ferromagnetic insulator EuS~\cite{Telkamp2026}, and with the chiral helimagnet Cr$_{1/3}$NbS$_2$~\cite{Beach2025}.

In the extrinsic JDE the barrier need not break any symmetry, because the device does.
A long-known example is the magnetic field of the bias current itself.
In 1967 Yamashita and Onodera showed that in cross-type junctions the self-field of the Josephson current shifts the field at which the critical current is largest by an amount that reverses with the current, and observed the resulting asymmetry of the current-voltage characteristic~\cite{Yamashita1967}.
The self-field also makes the interference pattern of window junctions asymmetric, by an amount set by the difference between the inductances of the two electrodes~\cite{Monaco2013}, and a self-field JDE is expected generally in planar junctions in a perpendicular field once the critical current is large enough~\cite{Rashidi2025}.
On its own the self-field asymmetry vanishes at the centre of the interference pattern, but any field offset that moves the centre away from zero applied field turns it into a zero-field diode.
Self-field diodes built on this principle use the stray field of a trapped Abrikosov vortex in planar Nb junctions, with a polarity switched by the bias configuration~\cite{Golod2022}, a control-line current in four-terminal junctions~\cite{Hovhannisyan2026}, and the magnetization of the barrier in long junctions with ferromagnetic GdN barriers and asymmetric current injection, with a polarity set by current pulses~\cite{Sharma2026}.
Offsets of this kind also arise from vortices pinned in the electrodes, which shift the interference pattern and make it hysteretic~\cite{Mangold2026}.
The bias current can also act through the kinetic inductance of thin superconducting leads, which turns it into a phase gradient along the junction that acts like a magnetic field~\cite{Chen2024,Ye2026}.
With such leads, feeding a planar junction from the same side or from opposite corners is no longer equivalent, which has led to the argument that nonreciprocity alone does not signal broken symmetry in the weak link~\cite{Ye2026}.

In a transport measurement the two classes can look alike.
In both, the critical-current patterns of the two polarities are displaced in opposite directions, and the asymmetry reverses with the current.
Once a remanent magnetization translates the pattern, an extrinsic diode operates at zero applied field with a polarity set by the magnetization, just like an intrinsic diode with a magnetic barrier.
A zero-field diode whose polarity follows the magnetic state of the barrier is therefore not by itself evidence of an intrinsic mechanism.
A dc measurement of a single junction characterizes the JDE but cannot probe the anomalous phase $\varphi_0$ directly.
Interferometers can measure it~\cite{Reinhardt2024,Jeon2026}, and Shapiro steps under microwave irradiation have been proposed to tell the two classes apart~\cite{Wang2025}, but both require additional circuitry or microwave excitation.
In junctions with the current perpendicular to the plane of the films, such as cross-type junctions, a simpler handle exists.
The self-field is produced by the current in the electrodes, so its direction, and with it the polarity of an extrinsic diode, is set by the path of the current to and from the junction and changes with the wiring, whereas an intrinsic diode depends only on the direction of the current through the barrier.

Here we show that remanent magnetization and self-field together mimic the zero-field JDE of a magnetic junction, and that the wiring of the current leads identifies the mimic.
We study cross-type junctions with Nb electrodes and a 12~nm thick Pd$_{89}$Ni$_{11}$ barrier.
PdNi is a weak ferromagnet widely used for Josephson $\pi$ junctions~\cite{Kontos2002,Khaire2009,Pham2022,BirgeSatchell2024}, and at this thickness it gives $\pi$ junctions with large critical current densities and has perpendicular magnetic anisotropy~\cite{Sapkota2026}.
The large critical current makes the self-field appreciable, and the anisotropy provides a clean reference state: an in-plane magnetization of a ferromagnetic barrier translates the interference pattern by its flux~\cite{Glick2017,BirgeSatchell2024}, whereas a barrier magnetized out of plane leaves the pattern centred~\cite{Satchell2021}.
The barrier as cooled therefore carries no net in-plane magnetization, and an in-plane remanent magnetization can be added afterwards with a magnetic field.
In the unmagnetized junction the self-field displaces the patterns of the two polarities to opposite sides of the centre.
Changing the wiring shows that the sign of the asymmetry follows the direction of the current in the electrode perpendicular to the field, as the self-field requires.
An in-plane remanent magnetization then translates the pattern by 3.2~mT and moves the asymmetry to zero applied field: the junction functions as a zero-field diode whose polarity is set by the direction of the magnetization.
The same change of wiring reverses the zero-field diode.
The magnetized junction thus mimics the zero-field JDE, and the wiring of the current leads provides a simple dc test that separates an extrinsic from an intrinsic diode effect in a single junction.

\section{Model}\label{sec:model}

The geometry is shown in Fig.~\ref{fig:model}(a).
Two Nb electrodes of width $a = 18$~\textmu m cross at right angles, with the circular junction of width $w = 3$~\textmu m at the crossing.
The bottom electrode E$_\parallel$ runs along the applied in-plane field $H$ and the top electrode E$_\perp$ across it.
A positive bias current $+I$ enters along one arm of E$_\parallel$, flows up through the barrier and leaves along one arm of E$_\perp$; over the junction area the sheet current of each electrode ramps between $I/a$ and zero as the current is transferred across the barrier.

The critical current of a circular junction in an in-plane field follows the Airy pattern~\cite{BaronePaterno},
\begin{equation}
  I_c = I_{\max} \left| \frac{2 J_1(\pi \Phi/\Phi_0)}{\pi \Phi/\Phi_0} \right| , \qquad \Phi = \mu_0 (H - H_c)\, w\, d_{\rm eff} ,
  \label{eq:airy}
\end{equation}
where $J_1$ is the Bessel function of the first kind, $\Phi_0 = h/2e$ is the flux quantum, $I_{\max}$ is the height of the pattern and $H_c$ its centre.
$\Phi$ is the flux through the junction: the field threads the barrier and penetrates the electrodes, so that the magnetic thickness $d_{\rm eff}$ is the barrier thickness plus the penetration of the field into the two electrodes~\cite{BaronePaterno}.
The first node of the pattern lies at $\Phi = 1.22\,\Phi_0$, i.e.\ at the field $\mu_0 H_1 = 1.22\,\Phi_0 / w d_{\rm eff}$ from the centre.
For an unmagnetized barrier an extrinsic $H_c$ can be introduced by the remanent field of the superconducting magnet used for the measurements; for a magnetized barrier an intrinsic $H_c$ can arise from the magnetization of the barrier itself, to which we return below.

The bias current produces a field of its own at the barrier~\cite{Yamashita1967}.
The sheet current of E$_\perp$, which flows at right angles to $H$, produces in the barrier a field along the field axis,
\begin{equation}
  B_{\rm self} = \kappa I ,
  \label{eq:Bself}
\end{equation}
which reverses with the current; for a uniform sheet current $I/a$, $\kappa = \mu_0/2a$, i.e.\ $35$~\textmu T/mA for $a = 18$~\textmu m.
In the wiring of Fig.~\ref{fig:model}(a), $B_{\rm self}$ is antiparallel to $H$ for $+I$; moving the current lead to the other arm of E$_\perp$ reverses it.
The sheet current of E$_\parallel$ produces a field of the same size perpendicular to $H$, which the applied field cannot compensate; it lowers $I_{\max}$ by a few per cent at the currents used here and is neglected.
With the self-field added to the applied field, the flux through the junction depends on the current, and the switching currents of the two polarities are the self-consistent solutions of
\begin{equation}
  I_c^{\pm} = I_{\max} \left| \frac{2 J_1(\pi \Phi^{\pm}/\Phi_0)}{\pi \Phi^{\pm}/\Phi_0} \right| , \qquad \Phi^{\pm} = \left[ \mu_0 (H - H_c) \mp \kappa I_c^{\pm} \right] w\, d_{\rm eff} ,
  \label{eq:model}
\end{equation}
which we solve numerically.
The measured switching current is the smallest positive solution, the first current at which the ramped bias reaches the field-reduced maximum supercurrent.

Figure~\ref{fig:model}(b) shows the two solutions of Eq.~(\ref{eq:model}) for $H_c = 0$ with the parameters of the junction studied below.
The self-field changes neither the height nor the width of the pattern; it skews the two lobes in opposite directions.
Because the displacement is proportional to the current that actually flows, $\kappa I_c^{\pm}$, it is largest at the maxima, which move to $H_c \pm \kappa I_{\max}/\mu_0$, and vanishes at the nodes; the two curves are mirror images of each other about $H_c$ and cross there.
The asymmetry
\begin{equation}
  \eta = \frac{I_c^{+} - I_c^{-}}{I_c^{+} + I_c^{-}}
  \label{eq:eta}
\end{equation}
is therefore antisymmetric about the centre, zero at $H_c$ and at the nodes, and largest where the pattern is steepest, between the maximum and the first node (inset of Fig.~\ref{fig:model}(b)); for the parameters of Fig.~\ref{fig:model}(b) it reaches 13\%.
The asymmetry grows with the height of the pattern and with the self-field coefficient, and its sign on a given side of the maximum is reversed by reversing the direction of the current in E$_\perp$ or by exchanging the two current leads at the source, which exchanges $I_c^{+}$ and $I_c^{-}$.

An in-plane remanent magnetization $M$ of a ferromagnetic barrier of thickness $d_F$ adds the flux $\mu_0 M d_F w$ to that of the applied field in Eq.~(\ref{eq:airy}) and translates the whole pattern~\cite{BirgeSatchell2024},
\begin{equation}
  H_c = -\,\frac{M d_F}{d_{\rm eff}} .
  \label{eq:Hc}
\end{equation}
$H_c$ is fixed by the remanent magnetization and reverses with it.
It is the same for both current polarities, so the translation on its own produces no difference between $I_c^{+}$ and $I_c^{-}$.
A multidomain barrier translates the pattern through its net in-plane magnetization only; the nonuniform remainder adds a position-dependent phase across the junction, which lowers the maximum and distorts the lobes~\cite{BaronePaterno}.

The two ingredients together give the zero-field diode effect: the self-field produces the asymmetry between the two polarities, and the magnetization, by translating the pattern, places that asymmetry at zero applied field.
$\eta(0)$ reverses with $M$ because $H_c$ does.

\section{Results}\label{sec:results}

\subsection{The unmagnetized junction}\label{sec:unmagnetized}

Figure~\ref{fig:virgin} shows the junction as cooled, before any magnetic field larger than the $\pm 10$~mT of the measurement itself has been applied.
Pd$_{89}$Ni$_{11}$ films of this thickness have perpendicular magnetic anisotropy: in our previous work we report that the out-of-plane hysteresis loop is square, while in-plane minor loops over the field range used in junction measurements show no remanence and no hysteresis~\cite{Sapkota2026}.
The as-cooled barrier therefore carries no net in-plane magnetization, and the $H_c$ we expect is the extrinsic one from the measurement magnet.

Figure~\ref{fig:virgin}(a) shows $I_c^{+}$ and $I_c^{-}$ against the in-plane field at 5~K.
To check for hysteresis and reproducibility, the field was swept from $+10$ to $-10$~mT and back, with two $I$--$V$ curves of each current polarity collected at every field on each pass.
The two sweep directions agree to 0.3~mA rms, within the symbol size, so each point in Fig.~\ref{fig:virgin}(a) is the median of the four curves, with their spread as the error bar.
The pattern has the form of Fig.~\ref{fig:model}(b) predicted by the model.
Its centre, $\mu_0 H_c = +0.44 \pm 0.02$~mT, is small, consistent with the remanent field of the superconducting magnet (a small intrinsic contribution of the barrier cannot be excluded, but neither is it required).

The two polarities are displaced in opposite directions: the $I_c^{+}$ pattern to positive and the $I_c^{-}$ pattern to negative field, as expected for the wiring of Fig.~\ref{fig:model}(a), and they cross at the centre.
The lines are Eq.~(\ref{eq:model}) fitted jointly to both polarities over the central lobe ($|\mu_0 (H - H_c)| < 6$~mT) with the four free parameters $I_{\max} = 20.5$~mA, $\mu_0 H_1 = 7.87$~mT, $\mu_0 H_c = +0.44$~mT and $\kappa = 36 \pm 2$~\textmu T/mA.
The fitted $\kappa$ compares to the estimate $\mu_0/2a = 35$~\textmu T/mA.
Outside the central lobe the pattern departs from the Airy form.
This could be due to a multidomain magnetic barrier in the as-cooled state, to the junction being large compared to the Josephson penetration depth ($w \approx 5 \lambda_J$ for the critical current density of this junction), or to both.
The uniformity of the side lobes does not affect the central argument in this work.

The asymmetry $\eta(H)$ of the same data, Fig.~\ref{fig:virgin}(b), is antisymmetric about $H_c$, as in the inset of Fig.~\ref{fig:model}(b).
It passes through zero at the centre $H_c$, and reaches $-11.4 \pm 0.2$\% at $-2.0$~mT and $+10.8 \pm 0.3$\% at $+2.5$~mT, on either side of the maximum.
Figures~\ref{fig:virgin}(c) to (e) show the raw $I$--$V$ curves of both polarities at the fields marked in (a), near these extremes and at the centre.
The switching to the resistive branch is sharp; $I_c^{+}$ and $I_c^{-}$ are obtained from a fit of each curve with the resistively shunted junction model, with the normal-state resistance held at 0.65~m$\Omega$ for every curve (lines), and are marked by the dotted lines.
The unmagnetized junction thus shows the full asymmetry of Fig.~\ref{fig:model}(b).

\subsection{Device geometry test of the self-field origin}\label{sec:origin}

So far we have compared our model with the data, which it describes.
However, an intrinsic diode effect arising from the barrier would give the same observation: a displacement of the $I_c^{+}$ and $I_c^{-}$ patterns to opposite sides of the centre.
It is therefore key to define a test that distinguishes an intrinsic from a self-field diode effect.
In the crossed-electrode geometry of our devices, Fig.~\ref{fig:model}(a), a simple device geometry test can be defined.
A reversal of the current polarity is insufficient: for an intrinsic diode effect as for the self-field, the two patterns exchange sides.
For the self-field effect it is the direction of the current in the electrode perpendicular to the applied field, E$_\perp$, that provides the test.
For an intrinsic diode effect it should not matter which arms of the electrodes carry the current, whereas reversing the direction of the current in E$_\perp$ reverses $B_{\rm self}$ and therefore exchanges the sides of the two patterns.
Reversing the current in E$_\parallel$ reverses only the self-field component perpendicular to $H$, which displaces neither pattern.

Figure~\ref{fig:origin} shows $I_c^{\pm}(H)$ of the unmagnetized junction in five lead configurations: (a) the reference wiring of Fig.~\ref{fig:model}(a), (b) the current lead moved to the other arm of E$_\perp$, (c) to the other arm of E$_\parallel$, (d) both, and (e) the two leads exchanged at the source.
The changes were made at the breakout box with the sample cold and without any field beyond the $\pm 10$~mT sweep, so that the magnetic state of the barrier and the remanent field of the magnet were the same throughout: the centres of the five patterns lie within 0.2~mT of each other, and their maxima within 1.2~mA.
In the reference wiring (a) the $I_c^{+}$ pattern lies on the positive-field side of the centre and the $I_c^{-}$ pattern on the negative side, as in Fig.~\ref{fig:virgin}(a).
With the current lead on the other arm of E$_\perp$ (b) the two patterns exchange sides, and $I_c^{-}$ is the larger critical current at positive field.
With the lead on the other arm of E$_\parallel$ (c) the patterns are as in (a), and with both arms changed (d) they are as in (b).
Exchanging the leads at the source (e) exchanges $I_c^{+}$ and $I_c^{-}$ and gives the mirror image of (a), as it must whatever the origin of the asymmetry; it also shows that the asymmetry is not an artefact of the current source or of the voltage detection.
Figure~\ref{fig:origin}(f) collects the results as the sign of the asymmetry on the positive-field side of the centre, positive when $I_c^{+} > I_c^{-}$: it reverses with the arm of E$_\perp$ and with the source polarity, and not with the arm of E$_\parallel$.
The measured row of the table is the self-field row, and the intrinsic row fails in (b) and (d).
The asymmetry of the unmagnetized junction is thus the self-field of its own bias current, and its sign is set by the wiring of E$_\perp$.

\subsection{Magnetized junctions and the zero-field diode effect}\label{sec:magnetized}

To place the asymmetry at zero applied field, the barrier needs a net in-plane remanent magnetization.
We magnetize it by warming the junction to 15~K, above the transition temperature of the Nb electrodes and below the Curie temperature of PdNi, applying $\pm 1.5$~T in the plane, returning the field to zero and cooling back to 5~K, where $I_c^{\pm}(H)$ is measured over the same $\pm 10$~mT range as before.
Figures~\ref{fig:magnetized}(a) and (b), measured in the reference wiring of Fig.~\ref{fig:model}(a), show the result after the magnetization procedure.
The state was reproduced in several repeats of the procedure in each direction of the magnetizing field, warming to 15~K and repeating the magnetization between them; the patterns shown combine the repeats.
In both states the pattern is translated: the field at which the two polarities cross, which we take as the centre $H_c$, lies at $\mu_0 H_c = -3.2$~mT after $+1.5$~T and at $+3.2$~mT after $-1.5$~T, the sign of Eq.~(\ref{eq:Hc}) for a magnetization along the magnetizing field, and the height of the pattern is unchanged within 1~mA.
Through Eq.~(\ref{eq:Hc}) it corresponds to a net in-plane remanent magnetization of 23~kA/m.
The translated pattern is not, however, an exact copy of the unmagnetized one.
Its central lobe is skewed about the centre.
The distortion is the mirror image in the two states and is reproduced in every repeat, so it is a property of the magnetized barrier.
We attribute it to the magnetization being multidomain: only its net in-plane component translates the pattern through Eq.~(\ref{eq:Hc}), and the nonuniform remainder adds a position-dependent phase across the junction that distorts the lobe.
In this wiring, as in Fig.~\ref{fig:virgin}(a), the $I_c^{+}$ pattern still lies on the positive-field side of the centre in both states.

Figure~\ref{fig:magnetized}(c) shows $\eta(H)$ of the two magnetized states.
In each state $\eta(H)$ is translated by the same field as the centre of the pattern, so that it no longer passes through zero at $H = 0$, and the two states are approximate mirror images of each other.
Figure~\ref{fig:magnetized}(d) shows the raw $I$--$V$ curves at $H = 0$, with $\eta = +10.1$\% and $-12.5$\%.
The junction mimics a diode at zero field, with a polarity set by the direction of the remanent magnetization.

Finally, we apply the lead-geometry test of Sec.~\ref{sec:origin} to the zero-field diode.
In each magnetized state we measured an $I$--$V$ curve of each polarity at zero applied field in each of the five lead configurations of Fig.~\ref{fig:origin}, switched at the breakout box with the sample cold, and recorded the sign of the zero-field asymmetry, positive when $I_c^{+} > I_c^{-}$.
Figures~\ref{fig:magnetized}(e) and (f) give the signs expected for the self-field and for an intrinsic diode, which are those of Fig.~\ref{fig:origin}(f) with the overall sign set by the direction of the magnetization, together with the observed signs.
In both states the observed sign follows the self-field row in every configuration: it reverses when the current lead is moved to the other arm of E$_\perp$ and when the source polarity is exchanged, and it is unchanged when the lead is moved to the other arm of E$_\parallel$.
An intrinsic diode would have kept its sign under the E$_\perp$ arm swap.
The zero-field diode is therefore wired into the device in the same way as the field-induced asymmetry of Sec.~\ref{sec:origin}.

\section{Discussion}\label{sec:discussion}

Although it is not an intrinsic diode effect, the magnetized junction functions as a zero-field Josephson diode with nonreciprocal critical currents.
Its direction is written by the field history, an in-plane field applied above the transition temperature of the electrodes, and is retained at zero field, so that the polarity is nonvolatile.
The junction is built from standard Nb and PdNi films and needs no exotic mechanism.

These properties are also the signatures by which an intrinsic zero-field JDE is usually identified: a nonreciprocal critical current at zero field, a sign written by the magnetization and retained, and an asymmetry that changes sign at the centre of the pattern (Fig.~\ref{fig:magnetized}).
The usual reversals do not distinguish the two either (Fig.~\ref{fig:origin}(f)): reversing the current exchanges $I_c^{+}$ and $I_c^{-}$ whatever the origin of the asymmetry, and the sign change of $\eta$ at $H_c$ matches that of an intrinsic diode driven by the field.
Only the lead geometry separates them, in applied field (Fig.~\ref{fig:origin}(f)) and at zero field (Fig.~\ref{fig:magnetized}(e), (f)) alike.
Moving the current lead to the other arm of E$_\perp$ reverses the self-field diode and would leave an intrinsic one unchanged, whereas moving it to the other arm of E$_\parallel$ changes neither.
The test needs a cross-type junction with two arms on each electrode, four terminals in all.
A claim of an intrinsic diode should therefore state the wiring and, where the geometry allows, show that the asymmetry is unchanged when the current lead is moved to the other arm of the electrode perpendicular to the field.

The size of the effect is set by the ratio of the self-field displacement of the maxima, $\kappa I_{\max}/\mu_0$, to the width of the pattern, $H_1$, which is about one tenth here.
In a cross-type junction like ours, this ratio grows as the cube of the junction width and in proportion to the critical current density and to the magnetic thickness, and falls as the electrodes are made wider.
It is appreciable here because the PdNi barrier combines a large critical current density with a junction a few micrometres wide.

The model also gives design rules.
In the model the asymmetry is largest where the pattern is steepest, so the ideal remanent shift, Eq.~(\ref{eq:Hc}), places the steepest part of the pattern at zero field.
The thickness of the barrier and its remanence are the knobs for this shift, although a thicker barrier also lowers the critical current density and with it the asymmetry.
The size of the asymmetry is set separately by the ratio above: a wider junction, a higher critical current density and electrodes no wider than needed all increase it.
Thin electrodes, or superconductors with a large kinetic inductance, increase it further, because the bias current then also produces a phase gradient along the junction through the kinetic inductance~\cite{Chen2024,Ye2026}; this is also why a small critical current does not by itself rule the mimic out.

The effect belongs with the other extrinsic diodes.
Like the vortex diode of Golod and Krasnov, in which the stray field of a trapped Abrikosov vortex provides the offset, it has memory~\cite{Golod2022}; here the direction is stored in the magnetization of the barrier instead.
The self-field diode of Rashidi et al.\ in planar junctions in a perpendicular field~\cite{Rashidi2025} is the same physics, a translation of the pattern by the self-field, without the memory, since the offset there is the applied field.

\section{Conclusions}\label{sec:conclusions}

We have shown that the self-field of the bias current and the remanent magnetization of the barrier together mimic the zero-field Josephson diode effect in Nb/PdNi/Nb junctions.
In the unmagnetized junction the self-field displaces the interference patterns of the two current polarities in opposite directions, by an amount set by the current in the electrode perpendicular to the field, and gives an asymmetry that is antisymmetric about the centre of the pattern.
An in-plane remanent magnetization translates the pattern and moves this asymmetry to zero applied field, where the junction mimics a nonvolatile diode whose polarity is set by the direction of the magnetization.
None of these observations requires an intrinsic diode effect originating from an exotic property of the barrier.

Reversing the current, the field or the magnetization gives the same result for the mimic as for an intrinsic diode.
The wiring of the current leads does not: moving the current lead to the other arm of the electrode perpendicular to the field reverses the self-field diode, in applied field and at zero field alike, whereas an intrinsic diode depends only on the direction of the current through the barrier.
This dc test needs only a cross-type junction with two arms on each electrode, and it is a simple check for any zero-field diode in a junction with the current perpendicular to the plane, especially where the critical current is large or the electrodes have a large kinetic inductance.
Conversely, the same two ingredients offer a simple route to a nonvolatile zero-field diode built from standard junction materials.

\section{Methods}\label{sec:methods}

\begin{figure}[p]
  \centering
  \includegraphics[width=\textwidth]{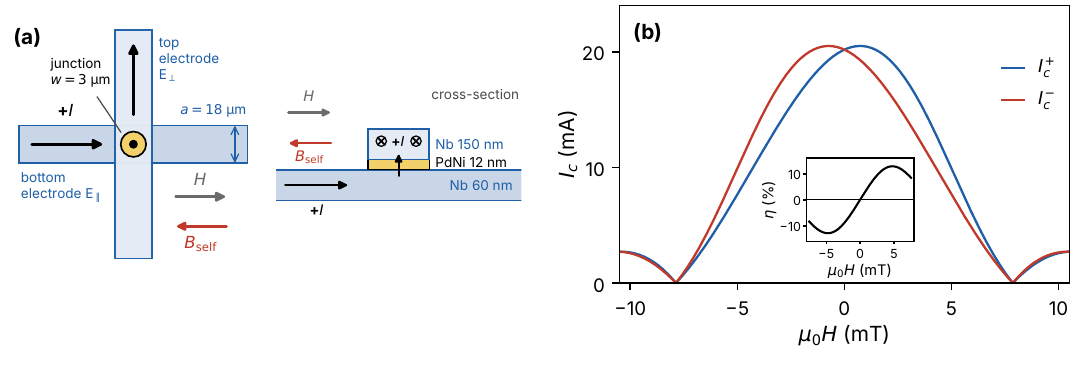}
  \caption{The self-field of the bias current in a cross-type junction.
  (a) The device from above and in cross-section.
  A positive current enters along the bottom electrode E$_\parallel$ (parallel to the applied field $H$), flows up through the circular junction and leaves along one arm of the top electrode E$_\perp$ (perpendicular to the applied field $H$).
  The current in E$_\perp$ produces at the barrier an in-plane field $B_{\rm self}$ along the field axis, antiparallel to $H$ for $+I$ in this wiring and reversing with the current.
  (b) The model of Eq.~(\ref{eq:model}) for the parameters fitted to the PdNi junction of Fig.~\ref{fig:virgin}.
  The $I_c^{+}$ pattern is displaced to positive and the $I_c^{-}$ pattern to negative field; the two coincide at $H = 0$ and at the nodes.
  Inset: $\eta(H)$ of the model, antisymmetric about the centre, zero there and at the nodes.}
  \label{fig:model}
\end{figure}

\bmhead{Device fabrication}
The junctions were fabricated by the process of our previous work~\cite{Sapkota2026}.
The Nb(60~nm)/Pd$_{89}$Ni$_{11}$(12~nm)/Pd(10~nm) stack was deposited by dc sputtering, without breaking vacuum, onto a Si substrate with a 100~nm thermal oxide in a system with a base pressure of $5 \times 10^{-8}$~Torr, through a lift-off mask that defines the bottom electrode E$_\parallel$ of width $a = 18$~\textmu m; the alloy was co-sputtered from Pd and Ni targets.
The circular junction of diameter $w = 3$~\textmu m was defined by a resist mask and Ar ion milling of the stack around it, then insulated with thermally evaporated SiO before the mask was lifted off.
The top Nb electrode (150~nm), of the same width and crossing the bottom one at the junction, was deposited through a second lift-off mask in a system with a base pressure of $2 \times 10^{-8}$~Torr after a few nm of the Pd cap had been removed by ion milling in situ.
Each electrode has an arm on either side of the junction, giving the four-terminal cross-type geometry of Fig.~\ref{fig:model}(a).

\bmhead{Measurement}
Transport measurements were made in a Quantum Design PPMS DynaCool with the sample on the multi-function probe, the plane of the junction parallel to the applied field and E$_\parallel$ along it, aligned by eye.
The bias current was supplied by the balanced current source module and the voltage across the junction read by the voltage measurement module of a Lake Shore M81-SSM synchronous source-measure system in dc mode.
Each $I$--$V$ curve is a ramp of the current from zero to $+23$ or $-23$~mA, from which $I_c^{+}$ or $I_c^{-}$ is extracted.
Interference patterns were recorded by stepping the field over $\pm 10$~mT in increments of 0.5~mT.
The temperature was 5~K, stable to within $\pm 5$~mK, and the thermometer on the probe showed no heating by the current ramps.
The lead configurations were set at the breakout box without otherwise changing the sample or its environment.

\begin{figure}[p]
  \centering
  \includegraphics[width=\textwidth]{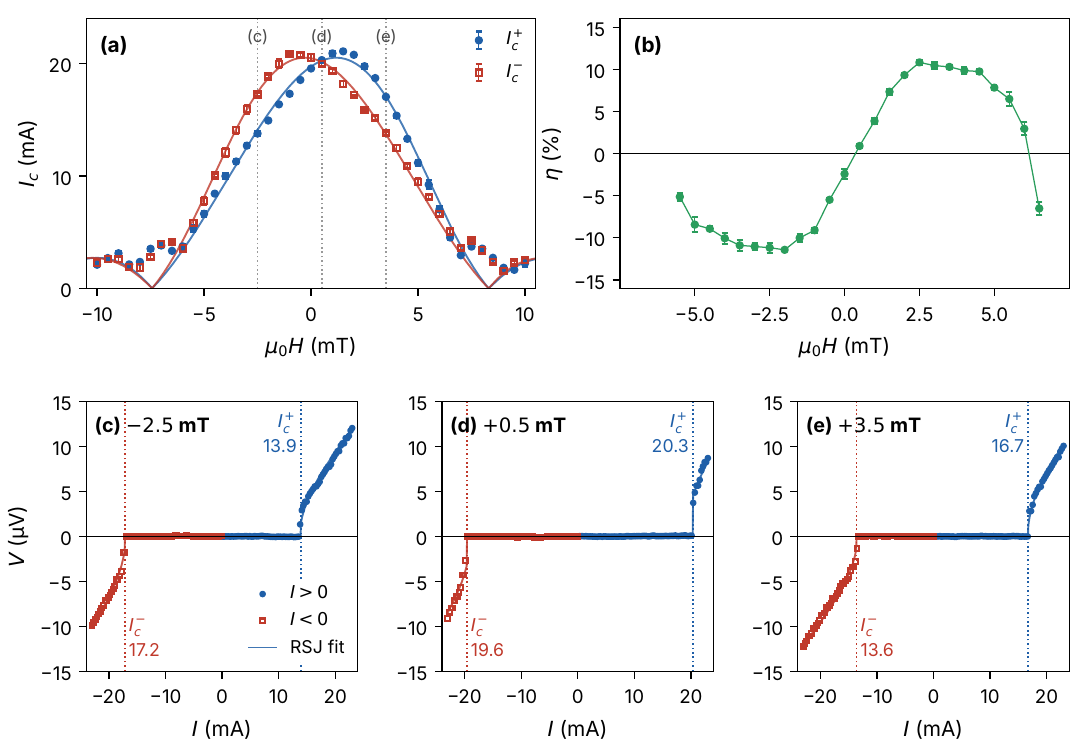}
  \caption{The unmagnetized Nb/PdNi(12 nm)/Nb junction at 5~K.
  (a) $I_c^{+}$ (filled circles) and $I_c^{-}$ (open squares) against the in-plane field, swept from $+10$ to $-10$~mT and back with two $I$--$V$ curves of each polarity at every field on each pass; each point is the median of the four curves, with their spread as error bar.
  Lines: Eq.~(\ref{eq:model}) fitted to the two patterns over the central lobe with four free parameters, $I_{\max} = 20.5$~mA, $\mu_0 H_1 = 7.87$~mT, $\mu_0 H_c = +0.44$~mT and $\kappa = 36 \pm 2$~\textmu T/mA.
  (b) $\eta(H)$ within the central lobe (error bars: standard error of the four repeats): antisymmetric about $H_c$, zero there, with extremes of $-11.4$\% at $-2.0$~mT and $+10.8$\% at $+2.5$~mT.
  (c) to (e) Raw $I$--$V$ curves of both polarities (points) at the three fields marked in (a), near the two extremes of $\eta$ and at the centre, with the fit of each curve with the resistively shunted junction (RSJ) model (lines).
  The dotted lines mark the fitted $I_c^{+}$ and $I_c^{-}$.}
  \label{fig:virgin}
\end{figure}

\begin{figure}[p]
  \centering
  \includegraphics[width=\textwidth]{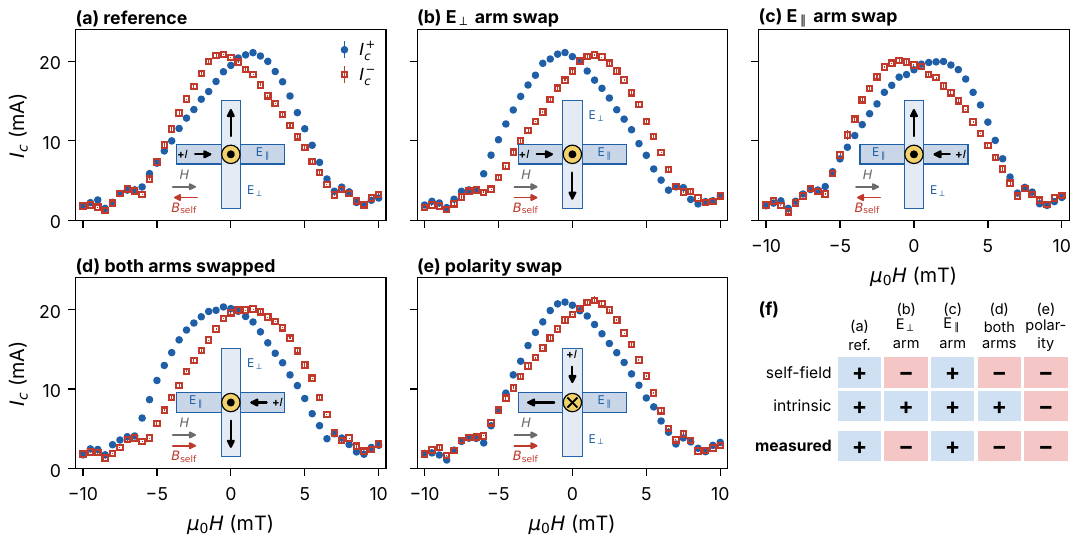}
  \caption{Device geometry test of the self-field origin: the patterns of the unmagnetized junction in five lead configurations, switched at the breakout box with the sample cold.
  Sketches: the device from above as in Fig.~\ref{fig:model}(a), with E$_\parallel$ (bottom electrode, along $H$), E$_\perp$ (top electrode, across $H$), the junction, the path of the positive current (black; dot: up through the barrier, cross: down), the applied field $H$ (grey) and the self-field of E$_\perp$ at the barrier, $B_{\rm self}$ (red).
  (a) Reference wiring of Fig.~\ref{fig:model}(a): $+I$ in along E$_\parallel$, up through the barrier, out along the top arm of E$_\perp$.
  (b) Current lead on the other arm of E$_\perp$: $B_{\rm self}$ reverses and the two patterns exchange sides.
  (c) Other arm of E$_\parallel$: no change.
  (d) Both arms.
  (e) Current leads exchanged at the source: $I_c^{+}$ and $I_c^{-}$ exchange.
  (f) The sign of the asymmetry on the positive-field side of the centre, positive when $I_c^{+} > I_c^{-}$, expected for a self-field and for an intrinsic effect under each operation, and the measured sign.}
  \label{fig:origin}
\end{figure}

\begin{figure}[p]
  \centering
  \includegraphics[width=\textwidth]{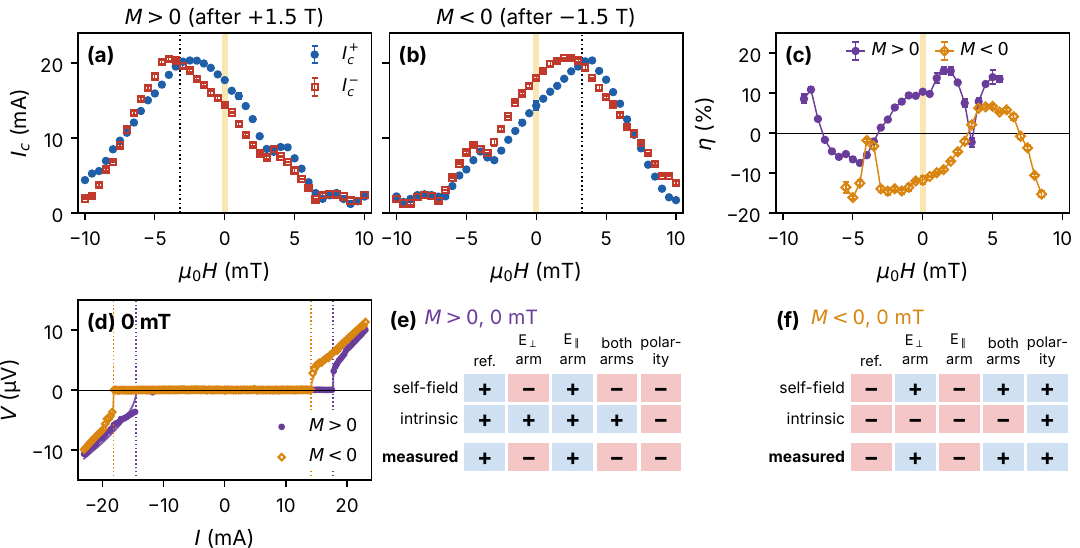}
  \caption{Magnetized junctions and the zero-field diode effect, 5~K.
  (a), (b) $I_c^{+}(H)$ and $I_c^{-}(H)$ after magnetizing the barrier at $+1.5$~T and at $-1.5$~T; each point is the median over the repeats, with their standard deviation as the error bar.
  The pattern is translated, with the crossing of the two polarities at $\mu_0 H_c = -3.2$ and $+3.2$~mT (dotted lines); zero field (shaded band) lies on the side of the pattern where the two polarities differ.
  (c) $\eta(H)$ of both states (error bars: standard error over the repeats), lines as guides to the eye.
  (d) $I$--$V$ curves at 0~mT in the two states (points) with their RSJ fits (lines); the dotted lines mark the fitted $I_c^{+}$ and $I_c^{-}$.
  The larger critical current is the positive one for $M > 0$ and the negative one for $M < 0$.
  (e), (f) The lead-geometry test of Fig.~\ref{fig:origin} at zero applied field in the two magnetized states: the sign of the zero-field asymmetry, positive when $I_c^{+} > I_c^{-}$, expected for the self-field and for an intrinsic diode in the five lead configurations of Fig.~\ref{fig:origin}(a) to (e), and the sign observed in one $I$--$V$ curve per configuration.}
  \label{fig:magnetized}
\end{figure}

\bmhead{Magnetic states}
The unmagnetized state of Figs.~\ref{fig:virgin} and \ref{fig:origin} was prepared by warming the junction to 250~K, well above the Curie temperature of PdNi (about 170~K~\cite{Sapkota2026}), and cooling in zero applied field to the 5~K measurement temperature.
To magnetize the barrier, the junction was warmed to 15~K, above the transition temperature of the Nb electrodes, a field of $+1.5$ or $-1.5$~T was applied in the plane along E$_\parallel$, the field was returned to zero and the junction was cooled back to 5~K before the pattern was measured.

\backmatter

\bmhead{Acknowledgements}

We thank Robert M. Klaes, Demet Korucu, Reza Loloee and Norman O. Birge for help with device fabrication and discussions.
We acknowledge fabrication assistance from Baokang Bi using the Keck Microfabrication Facility at Michigan State University.
This work was supported by the National Science Foundation under Award No.~2532911 and by new faculty startup funding made available by Texas State University.

\bmhead{Author contributions}

N.S. conceived the study, fabricated the devices, performed the measurements, analysed the data and wrote the manuscript.
A.S., P.S., K.L. and J.G. contributed to the measurements.
All authors reviewed the manuscript.

\bmhead{Data availability}

The data and analysis code that support the findings of this study are openly available in the Texas Data Repository at \url{https://doi.org/10.18738/T8/JPL6QI}~\cite{Satchell2026data}.

\bmhead{Competing interests}

The authors declare no competing interests.

\bibliography{PdNi_diode}

\clearpage

\clearpage

\clearpage

\clearpage

\end{document}